# Digital Technologies in the Context of University Transition and Disability: Theoretical and Empirical Advances

## *Edgar Pacheco*[1]

Since transition to higher education emerged as a research topic in the early 1970s, scholarly inquiry has focused on students without impairments and, what is more, little attention has been paid to the role of digital technologies. This article seeks to address this knowledge gap by looking at the university experiences of a group of first-year students with vision impairments from New Zealand, and the way they use digital tools, such as social media and mobile devices, to manage their transition-related challenges. The article summarises the findings from a longitudinal qualitative project which was methodologically informed by action research (AR). The article explores and discusses scholarly inquiry of transition to university and introduces a conceptual framework which includes five overlapping stages, the transition issues faced by the students and the roles played by digital technologies. The article updates and expands the theoretical understanding of transition to higher education and provides empirical evidence for practitioners to support the needs, inclusion, and participation of young people with disabilities in the tertiary setting.

[1] Victoria University of Wellington, School of Information Management. Wellington, New Zealand. e.pacheco1000@gmail.com

# 1      Introduction

There is an increasing, but still insufficient, body of knowledge looking at the role of digital technologies for supporting the needs of people with disabilities. Most of the available research about digital tools in the lives of people with disabilities, including the university experience, has centred on assistive technologies and their role of compensating for people's impairments (Pacheco et al., 2017). More recently, as new technologies have become more embedded in young people's everyday lives (Pacheco & Melhuish, 2018) research has explored different aspects of digital technologies in relation to people with disabilities, from identity development and self-representation (Thoreau, 2006) to the psychological and social impact of disability-specific online communities (Obst & Stafurik, 2010) as well as self-determination (Pacheco et al., 2019).

Despite the growing interest in both the transition to university of people with disabilities, and the implications of digital technologies in their everyday lives, research that links these two topics is to a large extent absent. This kind of research is currently more relevant considering the increasing number of students with disabilities attending tertiary education and the significant impact that transition has on the personal development of young people. Part of the limited literature has been written to provide practitioners with best practices (Bakken & Obiakor, 2008; Steere, Rose, & Cavaiuolo, 2007) or has scoped the transition needs of people with disabilities in general with no or little connection to the role of technology (see Kochhar-Bryant, Bassett, & Webb, 2009; Wagner, Newman, Cameto, Levine, & Marder, 2007). When digital technologies were related to tertiary education the focus was on access to these tools rather than their implications for the transition experience of the students (Burgstahler, 2003). In other cases, the relationship of digital technologies with transition to university has centred on the impact of assistive technologies in compensating for the impairments and/or helping students with disabilities adjust to the academic demands of the university setting.

With the rapid growth and increasing sophistication of digital technologies (e.g. social media, portable devices), researchers have started looking at their effects on the teaching and learning experience of university students. However, less attention has been paid to the impact of technology on transition to university. In particular, we still lack understanding of the way students with disabilities are using advanced new technologies to manage their transition. In other words, we do not know what roles digital technologies play in the transition experience of these students and how these tools are used to cope with different transition issues or factors apart from the academic demands of the university setting.

This article presents the findings of an AR study that updates the current understanding of transition to university – which to a large extent has ignored the role of digital technologies. It shows that students with vision impairments use and adapt digital technologies innovatively to manage transition issues or challenges and that, as a result, they also develop self-determination. In this sense, the role of new technologies is not limited to compensating for the vision impairment of the students. Digital technologies are also used for learning, collaboration and participation, among other roles, that allow the development of new skills and the empowerment of the students. The article develops Transition 2.0 as a new paradigm to study and support the transition experience of students experiencing disabilities. In addition to its scholarly relevance, Transition 2.0 also provides educators, disability service providers and policy makers with a new lens to understand the transition needs and experiences of students with disabilities.

The article is structured as follows. The next section discusses transition to university in the context of disability and summarises early approaches that in general have guided research about the topic. Then the research methodology followed for this study is described. Based on the findings, the subsequent section discusses the conventional view of transition which is called Transition 1.0. Then, the article introduces Transition 2.0 and develops its conceptual framework which includes the transition issues experienced by the students, the stages of Transition 2.0 and the different roles played by digital technologies for the transition journey.

## 2    Background

### 2.1 Transition to University and Disability

Transition to university involves a period of change that has an impact on all new tertiary students. It can be a more stressful and demanding experience for students with disabilities compared to the experience of their peers without disabilities. This is because disability makes transition a more challenging experience (Caton & Kagan, 2007; Kochhar-Bryant, Bassett, & Webb, 2009). The difficulties also increase when the students realise that the personalised support system they had in high school differs from the one at the university setting (Madaus, 2005). What is more, once at university, students with disabilities understand that they have to become independent learners. The evidence shows that the dropout rate of this group of students is almost double compared with students without disabilities and that they are less likely to graduate (Cobb, Sample, Alwell, & Johns, 2006; Bardin & Lewis, 2008; Thurlow, Sinclair, & Johnson, 2002) particularly because they are not prepared for the demands of the tertiary setting and/or have not received adequate transition preparation in high school (Hong, Ivy, Gonzalez, & Ehrensberger, 2007). Students with disabilities who manage to stay also spend more time studying at university before graduation (Caton & Kagan, 2007; Pacheco et al., 2019).

Managing transition to university can have a significant impact in other aspects of the life of students with disabilities. A successful transition experience can improve socialisation skills and expand personal social networks (Getzel & Wehman, 2005) or increase the chances of obtaining employment and gaining an income (Gilmore, Bose, & Hart, 2001; Wilson, Getzel, & Brown, 2000). Thus, in managing transition to university the students are not only more likely to succeed in their goal of obtaining a degree but also to become independent members of society.

### 2.2 Tinto's Theory of Student Departure

Attempts at explaining and theorising the transition to university date back more than forty years. Overall these approaches have focused on students' individual adjustment and adaptation to the university setting. However, as some have pointed out, the emphasis on adjustment and adaptation tends "to lead to research, policy and practice that are largely system driven and system serving" (Gale & Parker, 2011, p. 35). For instance, Spady's (1971) sociological model looks at the social factors that influence students' university experience such as family, previous educational background, academic potential, friendship support, grade performance, and social integration among others. Meanwhile, the psychological model (see Brower, 1992) centres on the personal characteristics that differentiate those students who persist from those who drop out of university and the students' need to adopt life tasks in order to succeed (e.g. academic achievement, social interaction, future goal development, autonomy, and time management). On the other hand, for the economic model (see Cabrera, Nora & Castañeda, 1992), students' financial concerns are the most important factor influencing their experiences in higher education. Financial issues increase anxieties and limit the amount of time and energy spent on academic activities and negatively affect academic performance and even students' social integration (Cabrera et al., 1993).

The most influential approach to the transition to university comes from the field of education through the work of Vincent Tinto and his theory of student departure. Tinto's theory outlines the complex process of students' integration and the crucial role that the social and academic systems of the tertiary institution play in their university experience. Both systems continually influence students and modify their original goals and commitments and thus lead them to either study completion or early departure (Tinto, 1975). The academic system is involved "almost entirely with the formal education of students" (Tinto, 1993, p. 106). It takes place not only in the classroom but also in other tertiary environments such as laboratories and involves various faculty and staff whose primary responsibility is the education of students (Tinto, 1993). On the other hand, the social system focuses on the many members of the tertiary institution, especially the students, and their social and intellectual needs. The social system is shaped by social interactions among students, faculty and staff that mainly happen outside the formal academic system of the tertiary setting, for instance, university halls, cafeterias and student clubs (Tinto, 1993).

According to Tinto (1993) the university experience involves three "stages of passage" that students "must typically pass in order to persist in college" (p. 94). Separation is the first stage. It is an isolating and stressful, if not a temporarily disorienting, experience. Students have to "disassociate themselves, in varying degrees, from membership in the past communities, most typically those associated with the local high school and place of residence" including their families (Tinto 1988, p. 95). The second stage, transition, refers to the period of passage between the old and the new and occurs during and after the stage of separation (Tinto, 1993, p. 97). Students experiencing transition "have not yet established the personal bonds which underlie community membership. As a result, they are neither bound strongly to the past, nor firmly tied to the future" (Tinto, 1988, p. 444). Transition is a critical period and the most challenging stage in students' university experience (Tinto, 1993). In the third stage, incorporation, the students are looking for integration and membership in the tertiary institution (Tinto, 1988). Social interactions are the primary means for achieving incorporation. Those who are unable to develop such interactions are likely to experience integration failure and its associated sense of isolation which could cause withdrawal (Tinto, 1988).

When Tinto developed and refined his theory, digital technologies such as the internet were not in the stage of commercial diffusion and were only used and accessed by very few people (Bell, 1999). Even computer-mediated networks were not widely used in the tertiary environment. Nowadays, digital technologies are present in students' everyday activities. Thus, there is a need to include these tools in the study of transition and to understand their role, as technological tools are not only changing the way education is delivered but also how students communicate and interact in the tertiary setting (Gatz & Hirt, 2000).

While Tinto's theory is a useful conceptual tool, its potential limitations have recently become more evident as a result of a changing world. The increasing number of students with disabilities, including those with vision impairments, attending university, poses a challenge for research, policy and practice. Thus, to understand the transition to university experience, research needs to consider the particular challenges and needs as well as the perceptions of students with disabilities and the way they make sense of their transition instead of employing a perspective that broadly aims at the adjustment of the students and their retention in, or departure from, the tertiary setting.

# 3 Research design

This study used a qualitative research approach. The methods and techniques of qualitative research allow us to understand, interpret and learn about the diverse and complex meanings of students with vision impairments of their transition to university. Instead of using schematic experimental procedures, the interactivity of the qualitative approach favoured a deeper understanding of students' perceptions and experiences (Berg, 2009). In addition, qualitative inquiry allows flexibility (Creswell, 2003). By using a qualitative approach, we were able to use an open research framework which was adjusted or refined when needed.

Action research (AR) was the chosen qualitative research method for the study. AR "aims to contribute both to the practical concerns of people in an immediate problematic situation and to the goals of social science by joint collaboration within a mutually acceptable ethical framework" (Rapoport, 1970, p. 499). AR is a rigorous research method that not only favours the use of a variety of qualitative techniques for data collection but more importantly supports iteration which makes the quality of the data and findings richer. Another reason for choosing AR was its problem-solving nature. While the purpose was to uncover new knowledge, some interventions were also conducted based on collaboration and the views of the participants to help them address transition challenges.

Data collection was a flexible process that took place throughout two AR cycles. The use of different techniques allowed the gathering of rich information at different stages of participants' transition experience. Research data was collected from observations, a researcher diary, online tools, focus groups and semi-structured interviews. Observations were particularly useful for obtaining data about the early stages of the transition experience of the participants when they were still prospective students or were in the first weeks of the academic trimester. A researcher diary complemented my observations. I used it to keep records, facilitate retrospective analysis, recall past thoughts and events, and evaluate the outcomes of my research (Borg, 2001). Data from online tools were collected via a Moodle-based website and a Facebook group page. They were also part of my AR interventions to provide transition support to the research participants. The website on the Moodle platform called "Goingtouni" helped to collect data about early transition issues. The Facebook group page was suggested by the participants. This data was collected in the form of online conversations, "likes" and the "seen by" feature on Facebook. While participants' online interaction via Facebook was mainly private, I managed to encourage some online group conversation that allowed me to expand my understanding of students' transition experience.

Three focus groups were conducted during the second cycle of the AR study because limited interaction and participation was obtained in the first cycle. While focus groups were a means for data collection, they were also part of the AR intervention to support the transition to university experience of the participants. The focus groups allowed the author to understand how the perceptions of the participants about their transition evolved through the first weeks of the academic trimester. Semi-structured interviews were also conducted after the participants completed their first trimester at university. In the interviews the participants evaluated their transition experience, and the way digital tools were used to manage transition challenges.

### *3.1 Research Participants*

The selection of the participants for this AR study was purposeful and focused on students with vision impairments from Victoria University of Wellington. As Patton (2002) suggests, working with "homogeneous samples" allows the description of "some particular subgroup in depth" (p. 235). Participants were undergraduate students, aged between 18 and 24 years old, first enrolled at Victoria University of Wellington in trimesters 1 and 2, 2012, and trimester 1, 2013. Of the 19 research participants, 17 were first-year students. The remaining 2 participants were senior students who took part in a pilot. Over a third of the participants came from Wellington and the rest from different cities and rural areas of New Zealand. The schooling background of the students was diverse. Some came from special education schools, boarding schools and public schools where they had received dedicated teaching support. Students coming from outside Wellington were living in university accommodation or flatting. Almost all the participants were school leavers and only a few of them had been working and living independently before arriving at university.

### *3.2 Data Analysis*

An inductive approach was used for data analysis. Inductive analysis is "making sense of data" (Lincoln & Guba, 1985, p. 202). It is an approach through which patterns, categories, and themes are built from the "bottom-up", by organising the data into increasingly more abstract units of information. Through adopting an inductive approach, the findings of the study are data-oriented. In other words, instead of confirming a hypothesis and theoretical pre-assumptions (Bryman, 2008), the concepts, categories and conceptual framework developed for the study were based on what the data was revealing. Transcripts were read and interviews were listened to several times to identify themes, to refine interpretations, and to allow new categories to emerge.

Data analysis started in parallel with data collection. It was an ongoing and iterative process that not only helped to refine interpretation of the findings but also to improve AR interventions and the transition support provided to the participants. It was ongoing because it took place along the different stages of the AR cycles and it was iterative because my interpretation of the data was refined several times as more data was collected and I reflected in more depth about their meaning. The following sections present and discuss the main findings of the study.

## 4        Transition 1.0

The analysis of transition has been dominated by a conventional view, which hereafter is called Transition 1.0. Transition 1.0 has focused on the personal need of the students to adjust to the university setting. It is seen as a period of psychological and academic changes for the students, who also must deal with an unknown university environment where there is no longer a dedicated and personalised support system available for them. The student not only feels alone and stressed but also has to rapidly learn the skills for independent learning. In Transition 1.0, students with disabilities are accommodated to fit into the demands of the tertiary setting, especially in regard to its academic responsibilities. Transition 1.0 centres on the normalisation of the students, so they are able to perform academically as *normal* students in order to obtain satisfactory grades. Thus, responses to smooth over the transition experience of the students have centred on compensating for and/or ameliorating the impact of their impairment through the provision of a range of specialised services and resources. These actions have also sought to promote equity and inclusion and reduce the physical and cultural barriers that prevent the students from engaging in and adjusting to university.

Transition 1.0 has included interventions to teach the students self-determination skills which will enable them to manage their transition experience. However, the scope of such interventions has only focused on instructing the students to learn the skills required for improving their academic performance (Fowler et al., 2007). In Transition 1.0, the students are still seen as passive recipients of support. Thus, self-determination interventions have mostly aimed at the students being able to function and/or adjust themselves according to the university demands. Independently of the support provided, Transition 1.0 is, to a large extent, a personal journey in which students with disabilities have to cope with transition challenges on their own and fit in at university.

In Transition 1.0, assistive technologies have played a primary role in compensating for the impairment of the students. These tools are used to support students' academic duties and performance. For example, the students are encouraged to use assistive technologies such as electronic Braille and screen magnification software to help them to read course material.

Transition 1.0 also involved the opportunities brought by the inception of the internet and the increased use of personal computers in the 1990s. For students with disabilities these kinds of digital tools have facilitated, for example, access to information and communication. The use of a personal email service, for instance, made it easier to be in contact with friends and relatives, overcoming issues of distance and time. Similarly, a personal computer adapted with other assistive technology allowed the student to enlarge the fonts or change the brightness on the screen, making course material readable. Despite the benefits, the technological developments of that period, particularly web interface and design, also increased the concerns regarding accessibility and usability. Some have speculated that digital tools were disabling (Goggin & Newell, 2003) and/or creating a disability divide for people with disabilities (Dobransky & Hargittai, 2006).

To sum up, Transition 1.0 centres on the changes experienced by the students at university at an individual level and the support provided to them to accommodate and/or compensate for the impact of their impairments. To a large extent, in Transition 1.0 the students make sense of their transition on their own, individually, and digital technologies are basically used to support impairment compensation and manage the academic challenges of the students with vision impairments. In other words, digital tools are fundamentally used to reinforce the transition experience as an individual journey.

## 5        Transition 2.0

Transition 2.0 represents a significant shift from the way transition to university has conventionally been seen by scholars and practitioners. The evidence from this study shows that students with vision impairments are using digital tools innovatively and creatively to cope with the challenges of their transition experience. Transition 2.0 does not mean a rupture from the conventional approach of Transition 1.0 but an evolution from it. The elements that characterise Transition 1.0 are still present in Transition 2.0. For example, in Transition 2.0 the students are still recipients of specialised support, indeed they still need it, and use assistive technologies to compensate for their vision impairments. The difference is that in Transition 2.0 the students have a more pro-active attitude. They understand that transition involves changes and challenges and they want to face them their way. They incorporate advanced technologies into their university experience so they are able to work, learn, collaborate and interact with their peers in order to manage their transition. As a result, transition to university becomes a collective experience in which the students are able to develop self-determination skills in their own fashion. Thus, Transition 2.0 can be defined as:

The personal and collective experience of the student with vision impairment of making sense of her or his transition to university by sharing, learning, interacting and collaborating via digital tools, especially social media and internet-enabled portable devices. In doing so, the student starts acquiring and/or developing the skills, attitudes and knowledge for self-determination that allow her or him to manage the challenges of university life and nurture her or his personal development as a young adult.

Transition 2.0 is driven by a behavioural and/or attitudinal change among students with vision impairments in relation to the way they see themselves at university and as young adults. The research findings show that the majority of participants also took the initiative in relation to their transition. Before the trimester started, they tried to foresee transition issues, arrange some support to manage their impairment and learn about how university works. They tried to make sense of their transition on their own terms. Such a can-do attitude did not mean that they did not make mistakes. They did so throughout the academic trimester, but they were also open to seeking support and advice, and readjusting their transition experience accordingly. In Transition 2.0, as one participant pointed out, students "do not want to be seen as somebody that is different". They, on the contrary, want to show that they are able to be part of the university and pursue their personal goals.

Not only do students with vision impairments feel capable of being at university, they also have no doubts that they are entitled to do so. In this respect, while in Transition 1.0 the students were thankful and believed that any university help they received was a favour, in Transition 2.0 attending university is perceived as a right and, thus, the students expect that their needs will be met by the tertiary institution. For example, the research findings show that a number of students talked with lecturers and course coordinators to let them know about their needs. One student was categorical that teaching staff "should understand my needs. I talked to them before every first lecture". Similarly, another student claimed that "lecturers should become familiar with this sort of thing [vision impairment]. They don't have to be reminded [about my needs]". A third student pointed out that "if I have the right things in place, if I have the things I need, I can be reasonably independent [as a student]". In other words, the students expected the university to share responsibility for their transition to university. There is clearly a behavioural and/or attitudinal change among the new generation of students with vision impairments that supports the claim for Transition 2.0. This move contrasts with Transition 1.0 in which the students were understood to be passive recipients of support who needed to be adjusted and normalised according to the demands of the tertiary setting. In contrast, Transition 2.0 is about students seeking to lead their transition and the university meeting the demands and personal needs of these students based on this new context.

On the other hand, Transition 2.0 differs from Transition 1.0 in that the students' transition is not only an individual experience but also a collective one. Certainly, each student with vision impairment deals with her or his transition in their own fashion. Their vision impairment, personality and background, for example, all have an impact on the way each of them experiences and manages transition challenges. However, in Transition 2.0, the transition is far from being an isolated journey. Making sense of it is also a collective endeavour in which the student learns along with their peers how to deal with different university challenges. There is extensive evidence from this research showing that students with vision impairments sought support from, worked and/or collaborated with their university peers, former high school friends and personal connections to manage different issues at different stages of their transition. In addition, once familiar with the university, these students were eager to advise and share their transition knowledge and experience with others. The implications of these findings suggest that universities should not ignore these actions/attitudes of the students that occur in parallel with the formal support provided by the tertiary settings.

In Transition 2.0 digital technologies are enablers used to make sense of transition challenges. The behaviour and attitudes of the students discussed in this section are supported by the use of diverse technological tools, not determined by them. Thus, the "2.0" in Transition 2.0 does not refer to an advanced version of the internet, known as Web 2.0, shaping students' university experience. In Transition 2.0 the students incorporate and adapt according to their personal needs a range of digital tools with which they are familiar. Web 2.0 along with assistive technologies and portable devices is part of these tools that support the actions of the students to manage their transition. For example, the research findings show that social media applications such as instant messaging and Skype complemented and/or supplemented face-to-face interaction and communication of the students with their families and friends. Similarly, the students also enhanced collaboration with their peers and their own learning via tools such as Facebook, YouTube and Blackboard. In other words, Transition 2.0 centres on the individual and collective experience of the students who used digital technologies to manage their transition.

Transition 2.0 also refers to a generation of young students with vision impairment which has grown up using technological tools as part of their daily lives. The students are, to a greater or lesser degree, competent technology users. This trend of technology competency has also been noticed by the Disability Services unit at Victoria University of Wellington. Some of its advisers agreed that "the students are very technology savvy" and that "even if we introduce new technology to them like a magnifying program that they haven't used before, because they've already got those skills in place, they can pick it up really easily". Indeed, the research findings show that a few students called themselves "ICT savvy". Transition 2.0 comprises, then, a generation of young students with vision impairments that does not limit itself to assistive tools. This generation is familiar with interactive and collaborative applications such as social media. They not only consume but also produce and share their own content online. They adapt tools such as digital cameras and voice recorders to cope with university challenges and enjoy the mobility provided by portable devices such as smartphones.

As previously mentioned, the findings of this research have uncovered a generation of students with vision impairment that want to be seen as unique and independent individuals. This suggests that transition can no longer be addressed in terms of making the students fit into university and/or only meeting academic demands. The behavioural changes and current use of digital technologies for transition show that in Transition 2.0 students with vision impairments are able to develop self-determination skills and that that has an impact on the way they manage their transition (Pacheco, Lips, & Yoong, 2019). Thus, Transition 2.0 should also be seen as an opportunity to support students' needs and aspirations of becoming self-determined young adults and not only well-adjusted students. Table 1 summarises the key differences between Transition 1.0 and Transition 2.0.

**Table 1**

*Transition 1.0 and Transition 2.0: Key differences*

| Transition 1.0 | Transition 2.0 |
|---|---|
| • It is the conventional view of transition to university.<br>• An individual journey for the student.<br>• In Transition 1.0 the student is expected to fit into the tertiary setting.<br>• The student is a passive receptor of support and "thankful" about it. She or he is still dependent on others.<br>• The student is concerned about disclosing her or his disability/impairment.<br>• Support focuses on "normalising" the student and ameliorating the impact of her or his impairment in the tertiary setting.<br>• There is a focus on teaching the student self-determination skills, mainly to perform well academically.<br>• Technology, namely assistive technologies, is used to compensate for student impairment at university. Other tools (e.g. the internet, email, desktop computers) are used to facilitate access to information and communication, to arrange disability support and assist learning. | • A paradigm shift about the way students with vision impairments experience their transition.<br>• Not only an individual journey but also a collective one, constructed in collaboration with peers.<br>• The student is aware that she or he is still expected to adjust to university but also requires the university to share responsibility for her or his transition.<br>• The student is pro-active about transition and seeks to manage it on her or his own terms and to learn about it by doing. She or he wants to show independence and self-assurance.<br>• The student is mainly open about disclosing her or his disability/impairment with their peers.<br>• The student does not ask but demands services and support to be available to her or him.<br>• The student uses and adapts a range of digital technologies (portable devices, social media) for her or his transition in addition to assistive technologies.<br>• Digital technologies are used to manage different transition issues (e.g. social connections, accommodation) and not only academic-related matters.<br>• In addition to access to information, communication and supporting learning, digital tools enable the student's participation, collaboration and interaction with and among their peers and the tertiary setting.<br>• The student develops self-determination which is supported by her or his use of digital technologies. |

*Note: adapted from Pacheco, Lips, and Yoong (2018).*

## 6      Towards a conceptual framework for Transition 2.0

Based on the research findings, a conceptual framework has been developed to understand transition to university for students with vision impairments (see Figure 1). The key idea in this framework is that the use and adaptation of digital technologies help the students to manage their transition, and, as result, to develop self-determination. The incorporation of the digital technologies and self-determination components not only represents the main contribution of the framework but also establishes a significant difference from the little prior research conducted in regard to this research topic.

Available research about transition to university and disability (see Belch, 2004; Duquette, 2000; Hodges & Keller, 1999; Wessel, Jones, Markle, & Westfall, 2009) has used Tinto's (1993) theory. According to Tinto (1993), students' backgrounds, as well as the level of academic and social integration with the university, determine their decision to stay in or leave the tertiary setting. However, Tinto's contribution does not consider the experience of students with disabilities. The focus of his theory is on the retention of the students to the tertiary setting. It does not cover the role that digital tools play in managing the transition experience and helping the students become self-determined young adults.

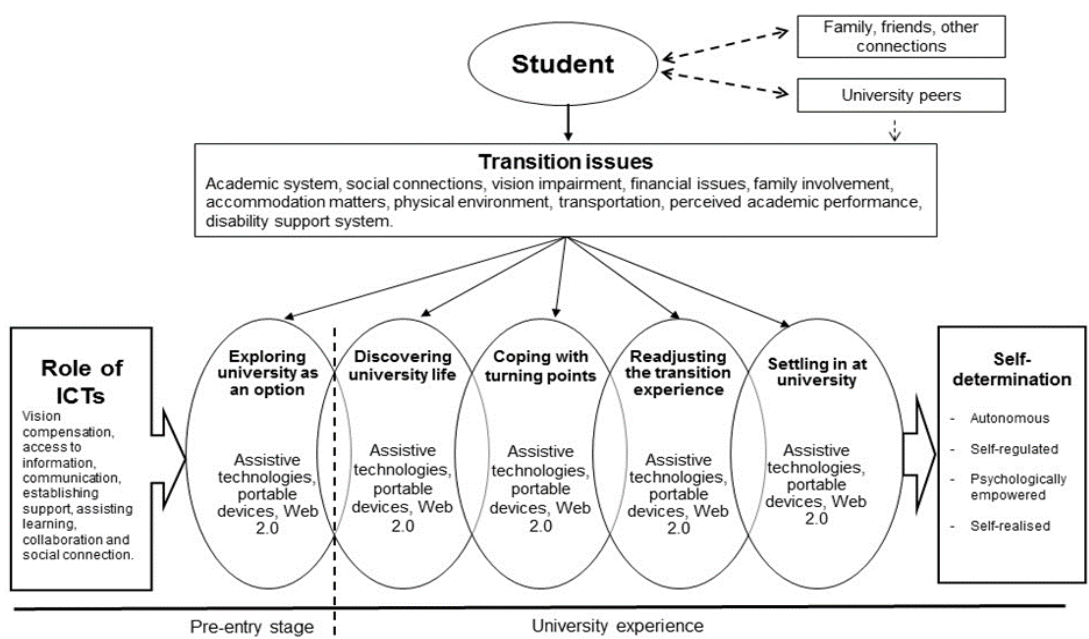

**Figure 1.** *Transition 2.0 for Students with Vision Impairments – conceptual framework (Pacheco, Lips, & Yoong, 2018, p.8)*

### 6.1 Transition Issues in Transition 2.0

In going through Transition 2.0, the student has to deal with different issues (Pacheco, Yoong, & Lips, 2020). These issues affect each student differently. As outlined by prior research, the study has also found that the academic system, social connections, transportation, family, accommodation, financial issues and vision impairment are issues experienced by the students. Although studies tended to see these issues as barriers, the participants in this research also perceived them as enabling their transition. Furthermore, this study unveils three transition issues not included in the literature: physical environment, perceived academic performance and support system (for further details see Pacheco et al., 2020).

Although all the students with vision impairments go through these transition issues, the way they experience them differs from one student to another. For instance, while the academic system was a critical issue for all the participants, some of them were more affected by specific aspects of it such as writing assignments or managing the large amount of course reading. On the other hand, the same transition issue, for example the social connections issue, was for some students a constraint but for others an incentive to stay at university. What is more, some issues which were considered at the beginning of the academic trimester as constraining the transition experience were later perceived as easing it. For example, some students who initially were unhappy and complained about their accommodation found later that it offered an appropriate environment for study.

More importantly, transition issues were found to be interconnected to the extent that managing one issue helped the students to cope with another one. A good example is the social connections issue. Students who were able to make new friends were also able to manage the academic system as they studied, supported and collaborated with their new network of friends. Finally, some transition issues were more lasting or recurrent than others. For instance, getting around the university's physical environment was an issue during the first weeks of the academic trimester until the students set daily routes from one lecture theatre to another. However, for some participants, financial constraints affected them throughout the entire trimester.

At first glance, the student faces transition issues on her or his own terms, individually. However, she or he also needs their university peers, existing friends and other networks or relationships to make sense of Transition 2.0. Thus, in the framework, Transition 2.0 is also described as a collective journey. Unlike Tinto, the framework is a holistic understanding of transition. Transition 2.0 is not circumscribed by the specific issues that occur within the tertiary setting (e.g. the academic and social systems of the university suggested in Tinto's). On the contrary, additional issues, such as family and financial constraints, including the impact of "external" social connections, are interrelated and also have an effect on the way the student experiences Transition 2.0.

### 6.2 The Stages of Transition 2.0

The framework includes the five stages the student with vision impairment goes through in Transition 2.0: exploring university as an option, discovering university life, coping with turning points, readjusting to the transition experience and settling in at university (see Pacheco, Lips, & Yoong, 2018). The overlapping ellipses in the framework illustrate that the stages are not independent and separate but ongoing and interconnected. The identification of these stages updates and expands previous descriptions of the transition process. A key feature is that they are dynamic and overlapping stages. The research findings show, for example, that while some students were in the midst of the discovering stage – still learning about the university's academic system and physical environment – they were also coping with the turning point of being unable to make new friends. These findings differ from previous research regarding disability and transition to university (see Duquette, 2000; Hadley, 2011) which has been guided by the work of Tinto (1993). While technology is absent from Tinto's stages, including other early approaches to the topic, research regarding disability and transition has limited digital technologies to the compensatory role of assistive technologies. The five stages of Transition 2.0 are summarised below.

### 6.2.1 Exploring university as an option

Exploring is a pre-entry stage. Students think and decide about tertiary studies when still in high school. University is seen as a pathway for personal development and a "passport" for independent life. Despite having priorities and set goals, students still need support.

Then they start seeking information and looking for specialised advice. Their main concerns are how to manage their impairment and learning about what university life looks like. They ask friends and other contacts about their university experiences. The participants are already users of a diverse range of assistive technologies for vision compensation. They also use the internet (e.g. the university website and YouTube) to find information about specialised support, accommodation and university entrance requirements. Personal email supports communication with university staff, particularly from the enrolment, accommodation and Disability Services units.

*6.2.2 Discovering university life*

The discovering stage refers to the first real encounter with university life. It starts from day one at university. The students feel anxious and lost and are concerned about having timely and adequate support in place (e.g. a note-taker in their lectures). The change from high school to university is perceived as an "overwhelming" experience. Their vision impairment is also perceived as a "hindrance". In this stage, some transition issues (physical environment, academic system, accommodation matters, and transportation among others) become more apparent. The increasing amount of course workload makes them feel stressed especially with reading course material. Digital technologies are used to deal with the first transition issues that emerge at the beginning of the academic trimester, for example the physical environment issue. The students, for example, used a map of the university in PDF format emailed to them or retrieved from the university's official website. Alternatively, they used the Google Maps application on their smartphones for campus navigation. In relation to the academic system issue, the students brought their laptops to the lectures, and in some cases, digital voice recorders and cameras, to support vision compensation and their learning experience. Similarly, they employed assistive technologies provided by the university such as a CCTV camera to manage reading tasks. The students also used social media and their smartphones to receive updates from their university halls of residence (accommodation matters) and search for information about bus timetables (transportation issue).

*6.2.3 Coping with turning points*

The participants identify and cope with turning points. Turning points are critical life events and/or experiences that make the student adopt changes and/or acquire new meanings about their transition. Turning points arise at any moment of the academic trimester and are not always caused by a negative event or experience. The main causes of turning points are events related to the academic system issue (e.g. writing assignments, the amount of course-related readings) and social connections issue (difficulties making new friends).

In this stage, digital technologies are an important medium used by students with vision impairments to manage turning points. They use Facebook (e.g. their personal profiles, course and research project pages) and the online forum on Blackboard to seek and share information. They collaborate and work with theirs peers via online platforms to manage a range of academic challenges (e.g. assignment writing). The social connection issue is managed by the use of tools such as Skype, Facebook, texting and instant messaging, which support communication and allow the students to keep in touch with their families, high school friends and university peers.

*6.2.4 Readjusting the transition experience*

After coping with turning points the students rethink and make changes that not only affect their transition but also their university experience as a whole. Changes are related to the academic system issue. In general, the participants decide on new directions or goals that include dropping papers, changing their enrolment status to part-time or changing their major in the next trimester. In some cases, they reaffirm previous goals. The readjusting stage reflects the commitment of the students in regard to their university-related long-term goals and personal development.

The students use digital technologies in order to make informed decisions. They refer to the university website to find information about administrative procedures for changing enrolment status and papers available in the next trimester. Then, again via the university website, they proceed to make the changes online. Social media (e.g. the Facebook group page for this research) was also used for receiving updates about next trimester deadlines. In addition to Facebook, other tools such as video calling software (e.g. Skype), texting and instant messaging were employed to seek feedback and advice from family members and other trusted social connections in real time.

*6.2.5 Settling in at university*

In the settling-in stage the students feel more familiar and in control of their transition experience. Although transition issues are still present, they are perceived as manageable. Overall the participants are more confident and secure, but they are also aware that they still need support. They perceive themselves as independent and self-determined young adults. Even those students who mention that they are still "finding their feet" have developed some self-determination skills. Settled-in students are predisposed to give advice and share the lessons from their transition experience. They recommend not being afraid about asking for support, planning ahead and getting support in advance.

The participants evaluated the role of digital technologies in their transition experience and concluded that these tools are "one of the biggest helps" for students with disabilities. From their experience, assistive technologies make it easier to manage their vision impairment. They advised new students to make sure that the required technology is in place and to get used to it before the trimester starts. The students also mentioned the different benefits of digital technologies for communication and support arrangement. In relation to social media tools, they concluded that these platforms have been "great" in supporting their transition to university, in particular when coping with the social connection issue.

**6.3 The Seven Roles of digital technologies**

The conceptual framework also includes a set of seven roles of technological tools in Transition 2.0. These roles, which expand the literature related to disability, transition and self-determination, are: enabling vision compensation, accessing information, facilitating communication, establishing and sustaining support, assisting learning, increasing collaboration, and achieving social connection and participation (Pacheco et al., 2017) (Pacheco et al., 2017). The following subsections develop on these roles.

*6.3.1 Enabling vision compensation*

The findings of this research support the idea of digital technologies helping students to compensate for their vision impairment. This argument is mainly related to assistive or adaptive technologies. Students with vision impairments unanimously outline the importance of these digital tools for ameliorating the impact of their impairment. These findings are in line with previous research based on the medical model of disability which has broadly highlighted the compensatory role of these tools in regard to the learning and studying experience of young people with disabilities.

The role of assistive technologies in vision compensation has a direct impact on the way students with vision impairments make sense of their transition experience. This can be observed when students have to manage some academic system-related tasks such as reading course material. As previously described, the students highlighted that their vision impairments are a "hindrance" for their transition experience. Having to read a large amount of printed course material through the academic trimester and being unable to see it properly quite often brought other effects such as feeling easily tired and losing concentration. They perceived that their reading pace was slower compared with students without disabilities. In other words, their vision impairments were clearly a disadvantage for their transition to university.

However, access to and use of different assistive tools offered students with vision impairments the opportunity to counteract these side effects and manage the impact of their impairment. From closed circuit television systems (CCTVs) to text enlargement software, the students were able to manipulate text according to their particular needs. A minor adjustment in the brightness and contrast of their monitors or laptop screens also made a difference and improved the readability of text. Other technological applications such as eBooks and PDF files offered similar opportunities to compensate for their vision impairment.

Vision compensation is enabled by the use of social media applications and portable devices as well. These kinds of digital technologies, used alone or in conjunction with assistive technologies, expand the abilities of these students to manage their impairments. For example, during the exploring and discovering stages, many participants preferred to use YouTube for accessing university-related information. For them, being able to listen to videos, instead of reading printed brochures and handbooks, was not only appealing but also useful in terms of avoiding tiredness and blurred vision. Portable devices with internet access, such as tablets and smartphones, have a similar impact. The data from this research show that participants took advantage of the built-in text enlargement and touchscreen features in their personal devices to compensate for their vision impairment.

*6.3.2 Accessing information*

The use of digital technologies makes it easier to access information for students with vision impairments. In their view, access to information is one of the most significant roles of digital tools in relation to their transition to university. On the one hand, digital technologies enable them to obtain information in a broader range of formats than just print. Information not only can be accessed digitally, for example, via eBooks and PowerPoint presentations; if printed, it can also be adjusted through assistive technologies, such as CCTVs. Moreover, students with vision impairments highly appreciated the fact that via technological tools they can search, retrieve and access a larger amount of information and from different sources.

Similarly, social media and portable devices appear to offer additional opportunities for accessing information in a dynamic and timely fashion. Participants in this research used social media and portable devices to access information. One of the participants highlighted the "convenience factor" of these tools. The research participants stressed that it was useful and appropriate to receive notifications straight to their mobile phones when information was posted on Facebook, the social media site of their choice. They were all familiar with that social networking site which they used, among other purposes, for receiving a variety of updates. Similarly, the Facebook group page set up for this research was quickly adopted as an additional source of information as updates went promptly to the participants' smartphones and were accessed anywhere at their convenience. In the same way, the university's official Facebook page and the course pages set up by some lecturers during the academic trimester were also regarded as useful sources of information for their transition.

Moreover, social media and portable devices are becoming primary ways to access information that allow students with vision impairments to manage different transition issues other than just the challenges related to the academic system. Along with conventional websites, social media were employed for accommodation matters. For instance, participants who lived in university halls of residence subscribed to the Facebook page of their accommodation in order to receive the latest updates directly to their personal profiles. They used these tools for transportation management as well. Some participants used their laptops regularly to look for bus timetables or, when at the bus stop, to find out the time the next bus would arrive.

The findings show that students with vision impairments are using portable devices with internet access to improve the way they search for information. For example, instead of using a printed bus timetable which was hard to read, many participants preferred to use their laptops or smartphones to plan their trip to university and other places. Through their devices, they were able to find information about bus and train services in real time. In the same way, getting to know how to move around the university campus was made easier for some participants when they accessed the Accessible Routes Maps via PDF files or Google Maps, a mobile web mapping service application set up by Disability Services. For other participants, online tools also made searching for information regarding products and services easier and eventually enabled them to do some online shopping and/or contact advocacy organisations such as the Blind Foundation. In summary, the use of a range of new technologies to access information helped the participants in coping with different transition challenges and becoming familiar with and more in control of their university life.

### 6.3.3 Facilitating communication

Digital technologies were an important communication medium for students with vision impairments. They indicated that digital technologies make it "a lot easier to communicate" because these tools offered them an array of channels that complement and/or supplement traditional forms of oral and writing communication. In practice, digital technologies supported participants' face-to-face communication with their close friends and relatives when there were barriers of time and distance. Prior research about disability (Bradley & Poppen, 2003; Seymour & Lupton, 2004) has highlighted the role of digital tools for improving communication; however, the findings of this research uncover its implications in the particular context of the transition experience of students with vision impairments.

For the participants, portable devices and social media applications were the preferred means of communication. All the participants had mobile phones with Wi-Fi connection, and they used their devices for texting and instant messaging with friends from high school and family members on a regular basis. Similarly, they used Facebook as a communication channel because it was "user-friendly" and made communication "quicker" and "better". Another reason why the participants used this platform was that all their friends were Facebook users. One participant revealed that she hardly ever uses the landline telephone at home to call her close contacts. She preferred Skype because she can see them on her laptop or smartphone. The attitude of this participant suggests that portable devices are diminishing the preference of students with vision impairments for "old" technology. Emailing was also used but to a lesser extent and for more "formal" communication such as contacting disability service providers or university staff. In summary, communication via portable devices and social media tools allowed the participants to overcome issues of distance and time and also had implications for other roles of technology such as support and collaboration, for instance.

### 6.3.4 Establishing and sustaining support

Digital technologies also play a role in establishing and sustaining support arrangements. Arranging adequate support was one of the primary concerns of the participants, especially during the few weeks before and after the start of the academic trimester. The majority of students were aware that they had moved away from the dedicated and specialised help received in high school. Therefore, they were concerned about having special course arrangements, assistive technology and other kinds of support in place at university. In general, digital technologies were an easier way to search for disability support information but also a convenient medium for contacting university service units, especially Disability Services and other service providers such as the Blind Foundation and to start arranging the required personal support and advice. For some participants, a phone call or an email was a preferable way to find out about and ask for transition assistance without disclosing too much about themselves and their disability. Once support was arranged, digital technologies enabled the participants to follow up with their Disability Advisers and promptly let them know if any other issue had arisen.

### 6.3.5 Assisting learning

The participants used a combination of digital technologies to enhance their learning experience, in particular the way they acquire knowledge and skills related to their chosen degrees. The majority of participants carried their laptops and smartphones, and in some cases tablets, to the lecture theatres. They downloaded and/or accessed via Blackboard the PowerPoint file of the lecture slides. Occasionally, they used the computer labs provided by the university when they needed to do some printing. Then, depending on their personal vision needs, they enlarged the content of the PowerPoint on their devices while following the presentation of the lecturer. Some participants, in addition, brought in their digital voice recorders and stored the recordings as MP3 files or other similar sound formats on their laptops so they could listen to them later. Alternatively, during lectures, a few participants used their smartphones to take pictures of the content on the whiteboard. These strategies of the participants to support their learning experience are similar to a growing tendency among university students in general who are using their personally owned portable devices to engage inside lectures (Gurung & Rutledge, 2014).

There are, however, some concerns that this trend, also known as Bring You Own Device (BYOD), may not have a positive impact on academic performance and, on the contrary, may cause student distraction as well as technical and teaching challenges for universities (Kobus, Rietveld, & van Ommeren, 2013; Traxler, 2013). Despite these worries, from the perspective of the participants in this research, bringing in their own portable devices with which they are familiar is clearly benefiting their transition experience at university.

Moreover, while formal learning takes place mainly in the physical settings of the university (lecture theatres, the library, study rooms, labs), students with vision impairment are also using social media and portable devices as a complementary environment for more informal and individualised learning. For instance, YouTube was used by some participants to support "big study". That is, independently of the quality of the information retrieved, these students used the video-streaming platform to search for further information and complement what was taught by the lecturer in class and/or obtain a better understanding of the essay topic they had to write about. In both cases, students with vision impairments adapted social media to respond to their personal learning needs.

These findings support recent scholarly discussion about the potential of social media for "learning on demand" (Punie, Cabrera, Bogdanowicz, Zinnbauer, & Navajas, 2005). In particular, the findings contribute to the growing interest in personal learning environments (Dabbagh & Kitsantas, 2012; Johnson, Adams, & Haywood, 2011; McLoughlin & Lee, 2007), which are seen as student-designed learning approaches that encompass different types of content – such as videos, apps, games and social media tools – chosen by a student to match his or her personal learning style and pace (Johnson et al., 2011, p. 8). In this respect, the participants do not limit themselves to textbooks and lectures in order to learn but also take advantage of alternative ways of learning via social media and internet-enabled portable devices. This complementarity between formal and informal learning offers valuable insights for understanding how this group of students manage transition challenges related to the academic system.

*6.3.6 Increasing collaboration*

The evidence in this research shows that digital technologies facilitate task collaboration. The majority of participants used social media and other interactive online applications set up on their portable devices to support online knowledge sharing. These applications provided the participants with an additional way to work together with their peers, especially in regard to academic tasks. This use of digital technologies did not replace but complemented conventional face-to-face forms of group work and study. Via these digital tools, the participants produced and/or shared diverse forms of content from comments to information. For example, some participants reported that they used Facebook to privately ask their peers for help and share ideas and information about academic matters. These tools allowed the participants to collaborate with each other outside of the university campus.

In addition, digital tools also supported online teamwork. The course pages set up by teaching staff on Facebook also enhanced cooperation and knowledge sharing with their lecture peers. Online collaborative work via the social networking site was especially useful for those participants who reported they had faced turning points related to the academic system. As these participants reported, social media allowed them to post questions, start group discussions and get feedback from other students who were also concerned and/or had some knowledge about particular academic tasks. Other web-based tools set up by the university also supported cooperation and knowledge sharing. The participants mentioned that Blackboard, the university's course management system, became not only a relevant source of information but also a tool for facilitating collaboration during different stages of their transition. Being able to collaborate and share knowledge about academic matters via digital technologies helped the participants in gaining confidence about their transition experience.

*6.3.7 Achieving social connection and participation*

The participants used social media to maintain existing relationships and to build new ones. Meeting new people at university was challenging for most participants, who felt isolated, especially at the beginning of the academic trimester. For them, making new friends was perceived as the way to fulfil their need for socialisation and to receive support and information regarding academic matters. One way to deal with the issue, while working on making new friends, was to turn to their "strong ties" (Putnam, 2000), in other words family and close friends. Meeting them face-to-face remained the participants' preferred type of social interaction, but the busy university life made meetings occasional if not difficult, especially for those who moved to Wellington to study. To counteract the barriers of distance and time, these students used social media to supplement online their limited physical social interactions with their strong ties. Applications such as Skype and Facebook, along with texting and emailing, were reported to be used regularly to cope with the lack of social connections, which in some cases became turning points for some participants.

Social media sites can also be used to support relationship building (Ellison, Steinfield, & Lampe, 2007) with "weak ties" (Putnam, 2000). The findings of this research confirm this claim in the context of the transition experience of students with vision impairments. Online interaction and participation via Facebook course pages set up by university teaching staff offered the participants the opportunity to share and receive valuable information from their peers, who were considered distant acquaintances rather than friends. The forum feature used in Blackboard also supported participants' social interaction with weak ties from their lectures. While this study did not show that social media and other interactive tools favoured the creation of online-only social connections, it supports the idea that these tools complemented and/or invigorated existing offline weak ties. When a repository website, called Goingtouni, was set up for this research, the participants were invited to participate in diverse online discussions. However, the students, who had not met each other previously, scarcely contributed to the discussion. In contrast, a second group of participants who took part in a number of group support meetings used their personal Facebook accounts to "catch up" with other members of the group. These meetings were the glue for the creation of new social connections among the participants and the use of social media complemented their need for social interaction and networking. In a few cases, these loose social connections became friendships.

The findings of this study suggest that social media and other interactive and portable tools helped students with vision impairments in managing the social connection issue. While face-to-face encounters were the preferred form of social interaction, the participants also used these tools to keep in touch with their existing connections and to build relationships with their emerging weak ties at university. In doing so, the participants coped with the feelings of isolation that concerned them from the beginning of their transition to university.

### 6.4 Towards self-determined students

As a corollary of actively using and adapting digital technologies for managing transition issues during the different stages of Transition 2.0, the student develops self-determination (Pacheco et al., 2019). In the framework, a self-determined student with vision impairment has the following characteristics: autonomy/independence, self-regulation, psychological empowerment and self-realisation. These four indicators are based on the work of Wehmeyer and Schwartz (1997) who have extensively researched self-determination in special education. The framework represents a major shift from previous studies. It does not restrict itself to the retention of the student at university and even less her or his adjustment and normalisation to fit into the tertiary setting. In contrast and based on the research findings, the framework suggests that, with the support of digital technologies, the student is able to develop the skills and behaviour to be independent and feel in control of her or his transition and, more importantly, to encourage her or his personal development – as a young adult. Thus, the framework challenges conventional scholarly views that tend to consider "successful" transition in terms of the student's grade performance, for example. As the research findings show, Transition 2.0 is an overarching experience in which the student does not manage academic matters only. In going through this critical life experience the student also seems to be learning and strengthening transferable abilities, skills and knowledge that would be crucial for their everyday life experience.

## 7 Conclusions

The findings of this study show that we need to rethink the transition to university. The literature describes transition to university as an individual experience. The student is seen as passive in relation to her or his transition and is meant to adapt to the demands of the university. In that context, digital technologies, especially assistive technologies, are used to ameliorate the impact of her or his impairment in the tertiary setting. This research, however, has found a different scenario. Nowadays, the student is pro-active and aware of the potential transition challenges posed by the impairment. The student perceives transition as a collective endeavour as well. She or he similarly adapts digital tools innovatively and uses them to participate, interact and collaborate with their peers in order to manage transition challenges. Because of the active use and permanent adaptation of these tools, the student is able to develop self-determination skills and uses these skills not only to cope with transition but also for her or his personal development as a young adult.

Considering this scenario, the findings reveal a paradigm shift to what could be called Transition 2.0. This new view of transition is based on the perceptions of the research participants and the way they constructed the meaning of their transition in interaction with their peers. Likewise, the current use of social media by the students does not define Transition 2.0. These tools are part of a new social and attitudinal context in which the student with vision impairment aims to be in charge of their transition. In this sense, social media and mobile devices and other digital technologies are enablers. The inception of Transition 2.0 in the scholarly analysis addresses a significant research gap in the study of transition to university and disability by incorporating in the analysis the implications of recent technological developments. It offers researchers, including policy makers, university teaching staff and service providers, with a new lens to understand and support the transition experience of young people with disabilities.

One of the main limitations of this study is its highly contextual nature. Although the findings are based on a rich set of data collected via different techniques and sources, they have to be taken with caution in regard to transferability. This research was conducted with the participation of students from Victoria University of Wellington. I am aware that it can be difficult to transfer the findings and conclusions of this inquiry to the context of other universities in New Zealand, or tertiary institutions in other countries. In addition, the results may neither be applicable to older students with vision impairments, nor students from other disability groups. However, they may be used as a lens when these particular contexts are researched. These limitations, however, open avenues for further study. Future research may also study how totally blind students manage transition to university. Another area of research could compare the transition experience of young and mature students with disabilities. In this study, some evidence from secondary sources of data revealed that older students are not benefiting from the use of digital technologies as their younger peers do. However, more research is needed as a large percentage of the New Zealand population with disabilities belong to the older age group. Finally, another avenue for further study would be research that includes other groups of students with disabilities. Although most research about transition has focused on students with learning disabilities, there is still a research gap regarding the impact of digital technologies on the transition experience of other groups of students with disabilities.

**Notes on contributor**

Edgar Pacheco holds a PhD in Information Systems from Victoria University of Wellington. Edgar is a research and policy analyst whose current work explores issues related to online safety including the risks and opportunities of digital technologies for children and adults, and vulnerable population groups. Edgar's research and scholarly published work also look at aspects related to human-computer interaction in the context of disability, transition, and higher education. https://orcid.org/0000-0003-4145-3244 @edgarpachecob1